# Multiple magnetic ordering phenomena in multiferroic o-HoMnO$_3$

Y. W. Windsor[1,2], M. Ramakrishnan[1], L. Rettig[1,2], A. Alberca[1], T. Lippert[3,4], C. W. Schneider[3], U. Staub[1]

[1] Swiss Light Source, Paul Scherrer Institute, 5232 Villigen PSI, Switzerland

[2] Department of Physical Chemistry, Fritz-Haber-Institute of the Max Planck Society, Faradayweg 4-6, Berlin 14915, Germany

[3] Laboratory for Multiscale Materials Experiments, Paul Scherrer Institute, 5232 Villigen PSI, Switzerland

[4] Department of Chemistry and Applied Biosciences, Laboratory of Inorganic Chemistry, ETH Zurich, Switzerland

## Abstract

Orthorhombic HoMnO$_3$ is a multiferroic in which Mn antiferromagnetic order induces ferroelectricity. A second transition occurs within the multiferroic phase, in which a strong enhancement of the ferroelectric polarization occurs concomitantly to antiferromagnetic ordering of Ho 4$f$ magnetic moments. Using the element selectivity of resonant X-ray diffraction, we study the magnetic order of the Mn 3d and Ho 4f moments. We explicitly show that the Mn magnetic order is affected by the Ho 4$f$ magnetic ordering transition. Based on the azimuthal dependence of the (0 $q$ 0) and (0 1-$q$ 0) magnetic reflections, we suggest that the Ho 4$f$ order is similar to that previously observed for Tb 4$f$ in TbMnO$_3$, which resembles an $ac$-cycloid. This is unlike the Mn order, which has already been shown to be different for the two materials. Using non-resonant diffraction, we show that the magnetically-induced ferroelectric lattice distortion is unaffected by the Ho ordering, suggesting a mechanism through which the Ho order affects polarization without affecting the lattice in the same manner as the Mn order.

## Introduction

Perovskite manganites of the o-$R$MnO$_3$ family (o- is short for orthorhombic, $R$ denotes a rare-earth lanthanide or Y) have been the subject of significant attention for over a decade, owing mostly to their magnetically induced ferroelectricity (type-II multiferroic behavior), which occurs for $R$ ions heavier (smaller) than Gd [1] (i.e. Tb-Lu and Y). All materials in this family exhibit antiferromagnetic (AF) orders of Mn moments that can be described by the same parametrized ordering wave vector **Q** = (0 $q$ 1), with different values of $q$. The Mn magnetic ordering pattern varies between these materials. For example, TbMnO$_3$ acquires a $bc$ cycloidal AF structure ($q \approx 0.27$) [2], while for LuMnO$_3$ an E-type AF structure is known to occur ($q = 0.5$) [3]. These AF orders induce a spontaneous ferroelectric polarization, ***P***. The magnitude of the ferroelectric polarization ***P*** induced by these AF orders depends on the ordering pattern: the AF cycloids are accompanied by $P \approx 500 \, \mu C/m^2$, while for the E-type AF a much larger value of $P \approx 5000 \mu C/m^2$ is expected [1]. In all cases the inverse Dzyaloshinskii-Moriya interaction contributes to $P$, but enlarged $P$ values are understood to arise due to symmetric exchange striction [4, 5].

Regardless of the exact Mn magnetic ordering motif, it is well-established that the ferroelectricity in these materials is induced by the Mn AF order, and not by that of the $R$ ions. An exception to this is the case of

o-HoMnO$_3$. Here an additional magnetic phase transition occurs well-within the multiferroic phase, below which P rises sharply by an order of magnitude. This behavior is illustrated in Figure 1. Neutron diffraction studies have associated this transition with an AF ordering of the Ho$^{3+}$ 4f moments [6]. Similar spontaneous R-ion magnetic order has been reported for other o-RMnO$_3$ materials, but at very low temperatures (e.g. for Tb, Dy and Tm this occurs at 7 K, 5 K and 4 K, respectively [7, 8] ). However, such a strong enhancement of P has not been reported for other R ions.

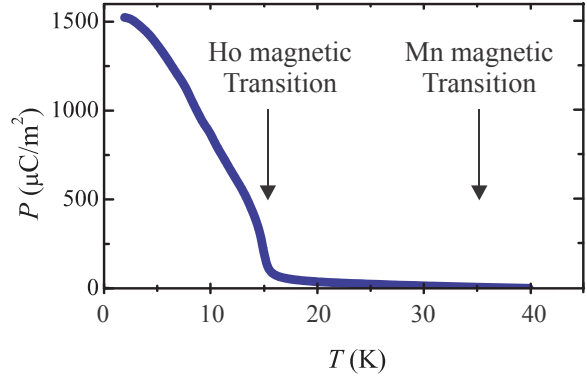

Figure 1 –polarization dependence of a crystal of HoMnO$_3$, adapted from Ref. [9]. Similar behavior is shown for a polycrystalline sample in Ref. [10].

An early study of o-HoMnO$_3$ suggested that Mn moments form an E-type AF order, implying a commensurate $q = 0.5$ [11]. Later studies indicated that an incommensurate AF order occurs with $q \approx 0.41$ [6, 9]. Nevertheless from the experimental phase diagram [12, 1] it is clear that HoMnO$_3$ is at the border between E-type AF and cycloidal AF structures, so variations in sample preparation could explain the difference in magnetic structure. Indeed more recent work by Shimamoto et al. has shown that epitaxial strain can "push" HoMnO$_3$ towards E-type AF behavior [13].

Upon cooling, the Mn moments in multiferroic o-RMnO$_3$ order in a sinusoidal antiferromagnetic (AF) structure below $T_N$. Upon further cooling they acquire a second magnetic component below $T_C$, producing the full Mn AF order, which induces a structural distortion and ferroelectricity (for the case of R = Ho, $T_N \approx 40$ K and $T_C \approx 35$ K). For heavy R ions the difference between $T_C$ and $T_N$ is only a few degrees K, so they are often not explicitly distinguished in literature. The Ho ordering transition, below which P rises sharply, occurs at a lower temperature of $T_{Ho1} \approx 14.5$ K [9]. The mechanism behind this enhancement is not known, but one could expect a further distortion of the lattice.

The study of HoMnO$_3$ and other heavy-R perovskite materials from this family has been limited due to the absence of single crystals. To date only one case of single crystalline HoMnO$_3$ has been reported [9]. In this work we study the magnetic and structural orders of o-HoMnO$_3$ using X-ray diffraction techniques on *thick* single-crystalline films. Using resonant X-ray diffraction (RXD) we observe the AF orders, and identify a strong coupling between the Mn 3d and the Ho 4f moments, specifically around the 14.5 K transition. Non-resonant diffraction is used to follow symmetry-forbidden Bragg reflections, the appearance of which herald the lowering of crystal symmetry that is associated with the occurrence of ferroelectricity. RXD is also used to provide a detailed account of the evolution of the Ho moments, which are shown to be strongly affected by three magnetic transitions.

## Experiment

o-HoMnO$_3$ films were grown on NdGaO$_3$ [010] substrates (Crystec Co. Ltd.) by pulsed laser deposition. Pulsed beams from a KrF excimer laser ($\lambda$ = 248 nm, $\tau$ = 25 ns) were imaged onto a stoichiometric hexagonal HoMnO$_3$ ceramic target with a fluence of 2.3 – 2.7 J cm$^{-2}$ at a repetition rate of 2 Hz in a N$_2$O partial pressure of 0.30 mbar. The target-substrate distance was fixed at 37 mm and the substrate was heated to 780°C during the film growth. The crystal structure and quality of the as-deposited films were investigated using Cu K$\alpha$ X-ray diffraction. Since thin films are known to acquire different magnetic order than bulk samples [13], all results are from thick films (120 nm, 240 nm) that exhibit the magnetic transitions described in the following.

Resonant X-ray diffraction experiments (RXD) were conducted using the RESOXS UHV diffraction end station [14] at the SIM beam line [15] of the Swiss Light Source (SLS). Photon energies used correspond to the Mn $L_{2,3}$ absorption edges (2p→3d, ~643eV and ~652eV) and to the Ho $M_{4,5}$ absorption edges (3d→4f, ~1390eV and ~1350eV). The large resonant enhancement of magnetic diffraction at these edges renders the collected signals solely sensitive to the resonant ion. Linearly polarized incident light was used, with either $\pi$ or $\sigma$ polarization (electric field in the scattering plane, or perpendicular to it, respectively). Scattered intensities were collected using an IRD AXUV100 photodiode. Azimuthal rotation of the sample around the momentum transfer vector $Q$ was conducted using a mechanical arm to a precision of ±3° (see experimental geometry in Figure 2). We define the azimuthal angle $\Psi$ as zero when the crystallographic $\hat{a}$ axis is in the scattering plane. Non-resonant X-ray diffraction (XRD) was conducted using a 3+2 diffractometer at the Surface Diffraction end station of the Materials Science beam line of the SLS [16]. The incoming photons had an energy of 10 keV, and were linearly polarized. Diffracted intensities were collected using a Pilatus 100K detector [17] mounted on the detector arm. In both experiments, samples were mounted on the cold head of a Janis ST-400 flow cryostat and cooled to temperatures as low as 9 K.

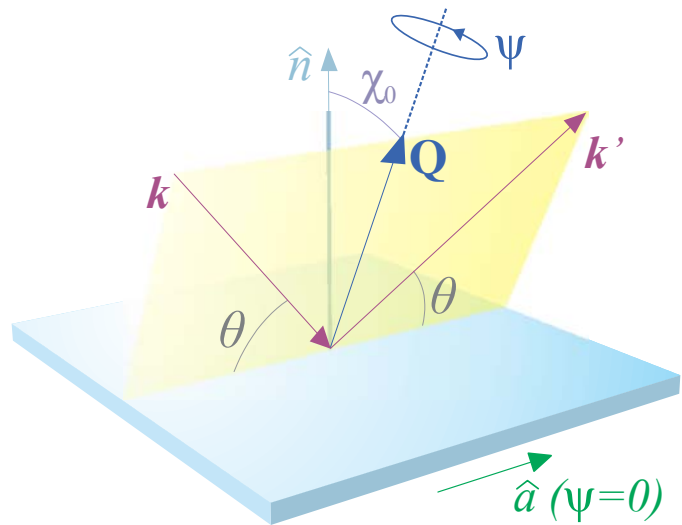

Figure 2 – experimental geometry of the RXD experiment. The scattering plane contains the momentum transfer $Q$ and the directions of the incoming and outgoing beam, $\vec{k}$ and $\vec{k}'$, respectively. $\theta$ is the Bragg angle and $\chi_0$ is the angle between $Q$ and the film surface normal $\hat{n}$. $\Psi$ is the angle of azimuthal sample rotation around the $Q$ direction.

## Results

I. RXD study of Magnetic Order

Figure 3a presents scans taken along the [010] direction at 10 K, using photon energies near the Mn $L_3$ and the Ho $M_5$ edges. In both cases a magnetic reflection is observed at (0 $q$ 0), with $q \approx 0.41$ (r.l.u.). This value coincides with that found in bulk studies, confirming that our thick films are bulk-like, and not significantly strained. An additional (0 1-$q$ 0) magnetic reflection is observed at the Ho edge, which is absent at the Mn edge. A (0 1 0) reflection due to Tempelton-Tempelton scattering is also observed at resonance (not shown), which is related to 4$f$ quadrupole ordering, and also persists in the paramagnetic phase (T > $T_N$). This is analogous to the (1 0 0) reflection observed for $o$-TmMnO$_3$ [18]. The inset presents the energy dependence of the (0 $q$ 0) reflection around the Mn $L$ edges at 9 and 18 K (above and below $T_{Ho,1}$ = 14.5 K). The data are normalized at the peaks of the $L_3$ edge, highlighting a shift in spectral weight between the two edges (see colored region at the $L_2$).

Figures 3b and 3c present the temperature dependence of these signals, measured at the Mn $L_3$ and Ho $M_5$ edges, respectively. For easy comparison, the data were collected in the same experiment and with the same geometry, using $\pi$ polarized light at $\Psi = 0°$. The (0 $q$ 0) reflection appears at both edges below $T_N \approx 39K$. The $T_{Ho1}$ transition is marked by a dashed red line. Clear signatures of this transition are observed at both resonances: at the Mn edge a dip in the (0 $q$ 0) intensity occurs, and upon further cooling this reflection's intensity at the Ho edge rapidly increases. Most importantly, below $T_{Ho1}$ the (0 1-$q$ 0) reflection appears at the Ho edge. To better quantify these observations, Figure 4a presents the integrated intensity of these reflections as functions of temperature. The Mn integrated intensity increases rapidly upon cooling below $T_N$, but flattens out below the Ho transitions. The Ho integrated intensities grow upon cooling below these transitions.

The higher energy of the Ho $M_5$ edge enlarges the Ewald sphere, which permits access to additional magnetic reflections. Figure 4b presents the integrated intensity of the (0 $q$ 1) and (0 1-$q$ 1) magnetic reflections, also measured using $\pi$ polarized light at $\Psi = 0°$. Both reflections intensify rapidly below $T_{Ho1}$. However, while the (0 $q$ 1) reflection appears below $T_N$ = 39 K, the (0 1-$q$ 1) appears below a third magnetic transition: $T_{Ho2}$ = 23 K. This transition was observed in Ref. [11], but with $q = ½$. At 9 K the integrated intensity ratio (0 1-$q$ 1) / (0 $q$ 1) is 0.48 or 0.66 for $\pi$ and $\sigma$ incoming light, respectively.

Figure 4c presents the temperature dependence of $q$, obtained from the position of the (0 $q$ 0) reflection in Fig. 3b and 3c. $q$ varies smoothly with temperature, and peaks around $T_{Ho1}$. The $q$ values from the Ho and Mn datasets closely follow each other, demonstrating that the Ho and Mn moments are coupled throughout the observed temperature range. The apparent shift between them is due to the difference in refractive index between the two energies, as similarly reported for $o$-TmMnO$_3$ in Ref. [19]. This wavevector relation has been noted as an indicator for stronger $R$-Mn coupling than in $o$-TbMnO$_3$ and $o$-DyMnO$_3$. [20]. Figure 4d presents the correlation length of (0 $q$ 0) at the Ho and Mn edges, defined as 2/FWHM of the reflection. A change in trend is observed at both edges at $T_{Ho1}$ and slightly below $T_N$. The changes in the Mn data around $T_{Ho1}$ occur ~1.5 K lower than the changes in the Ho data. This systematic difference is not clear around $T_N$. The correlations lengths of the off-specular reflections are also shown. These do not follow the same trend as (0 $q$ 0), and are likely limited by the reduced probe depth of these reflections.

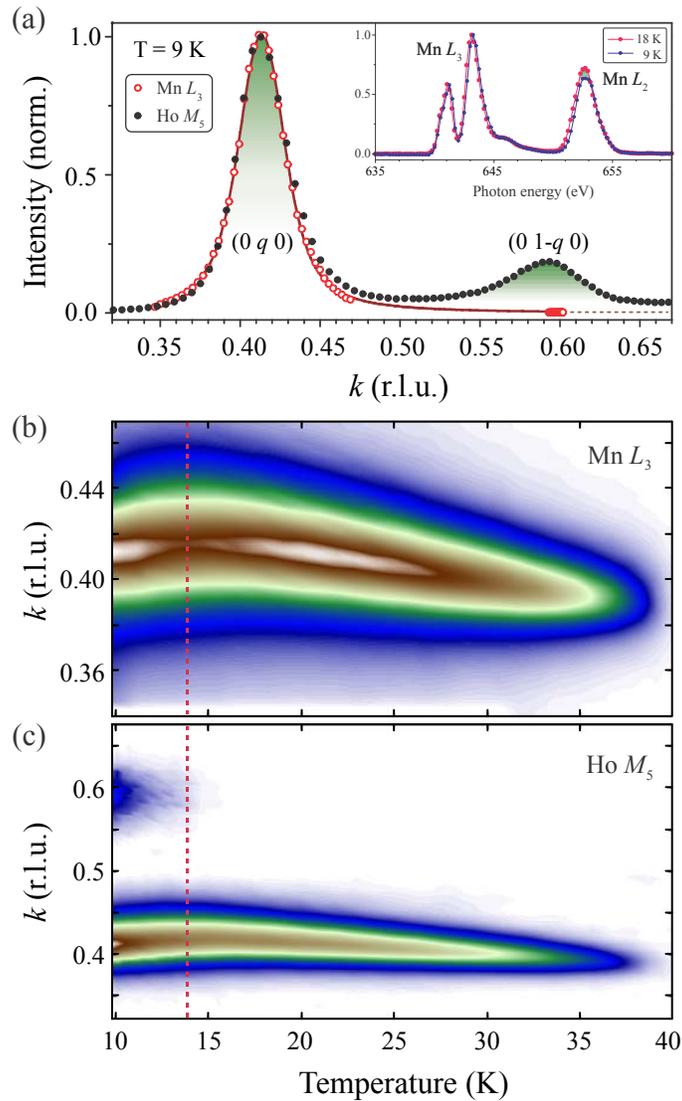

Figure 3 - (a) Scans along (0 $k$ 0) taken at 9 K using photons energies corresponding to the Mn $L_3$ and the Ho $M_5$ edges (642.5 eV and 1349.75 eV, respectively). The (0 $q$ 0) reflection is observed at both energies, and the (0 1-$q$ 0) is observed only for Ho. A dashed line marks the Mn curve beyond the Ewald sphere. Inset: energy dependence of the (0 $q$ 0) reflection at the Mn $L$ edges at 10K and at 18K (above and below $T_{Ho1}$), with intensities normalized to the peak of the $L_3$. (b) and (c) present a temperature dependence of scans along (0 $k$ 0), measured at the Mn $L_3$ and Ho $M_5$ edges, respectively. A dashed line is drawn at $T_{Ho1}$. All data were taken with $\pi$ polarized light and $\Psi = 0°$. The reflectivity background was removed for clarity.

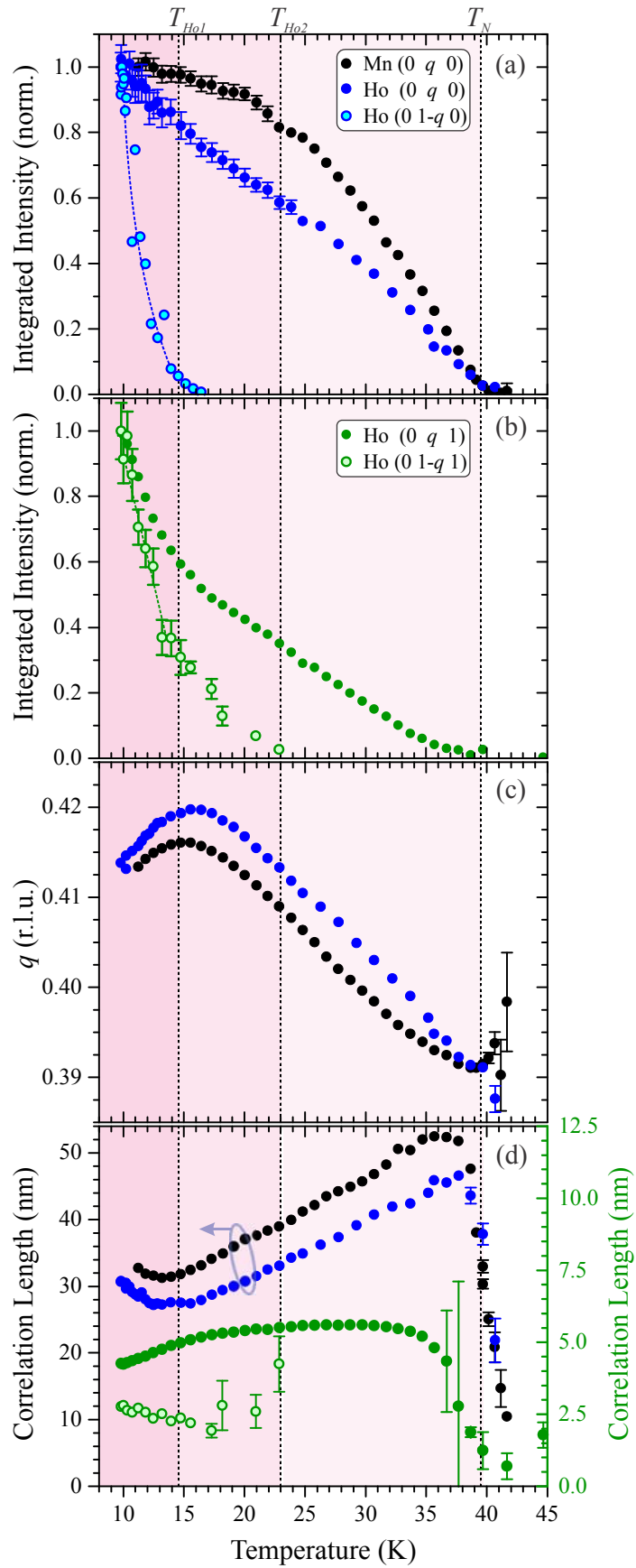

Figure 4 – Temperature dependences. (a) The integrated intensity from specular magnetic reflections measured at the Mn and Ho edges. (b) The integrated intensity from off-specular the reflections observed only at the Ho $M_5$ edge (1348 eV). (c) peak position of the (0 $q$ 0) reflection measured at both edges. (d) correlation lengths calculated from the widths of the magnetic reflections. Values for the off-spectral reflections are given on the right hand axis. All data were collected with $\pi$ polarized incoming light at $\Psi = 0°$.

The azimuthal angle dependence of the specular reflections' intensity was also collected. Figures 5a and 5b present the (0 q 0) reflection at the Mn and Ho edges, respectively, using π and σ incoming light at 18 K (between $T_{Ho1}$ and $T_{Ho2}$). Data are normalized to the sum ($I_\pi + I_\sigma$) to eliminate extrinsic artifacts. The solid lines represent calculations of the expected signal when only c axis moments contribute to it (detailed in previous works [21, 19]). This is in agreement with both the Mn and the Ho data. However, due to the high accuracy in the Ho data, deviations are observed. This is in contrast to the case of $TmMnO_3$, in which an excellent agreement was found both for the Mn and the Tm edges [19]. The deviation suggests an additional contribution to intensity at the Ho edge. Because the Ho ions are in the 4c position of the unit cell, such a contribution could indeed occur if Ho moment components exist that are perpendicular to the c axis (see appendix). Lastly, Figure 5c presents the azimuthal dependence of (0 1-q 0) at ~10 K. Large uncertainties are apparent because this reflection appears only below $T_{Ho1}$, and is therefore weak, as this is near the lower limit of the experimental cooling range. Nevertheless, the azimuthal pattern clearly exhibits a π/2 phase shift compared to the (0 q 0) intensities in Fig. 5a and 5b.

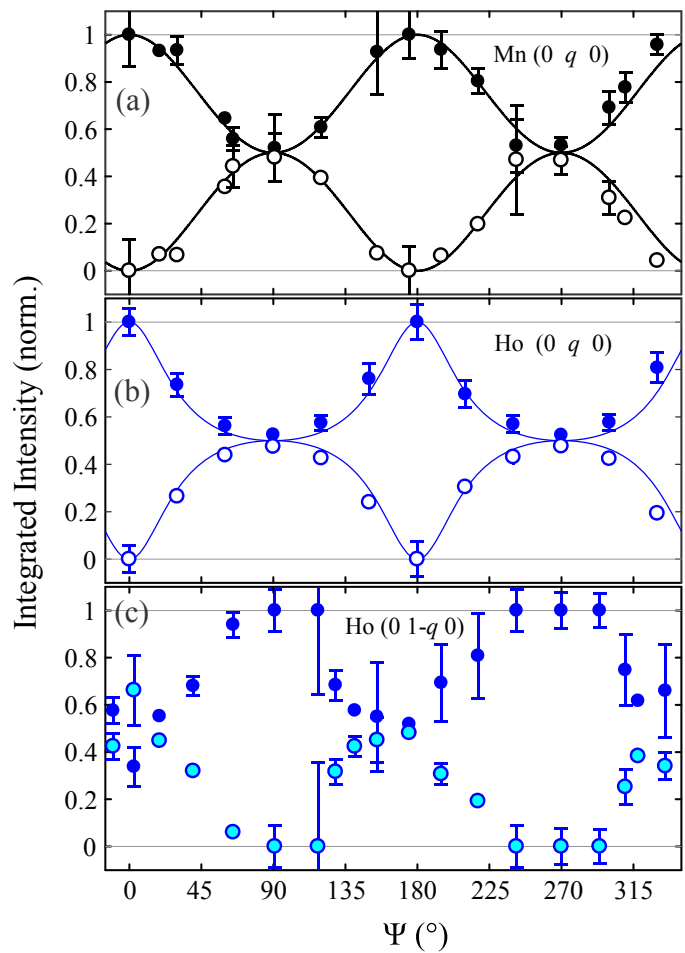

Figure 5 – Azimuthal angle dependence of specular magnetic reflections: (a) the (0 q 0) at the Mn $L_3$ edge (18 K); (b) the (0 q 0) at the Ho $M_5$ edge (18 K; very similar results we obtained at ~10 K); (c) the (0 1-q 0) at the Ho $M_5$ edge (~10 K). Filled and open markers represent π and σ polarized incoming light, respectively. All data are normalized to the sum of π and σ intensities. Solid lines are calculations of c axis magnetic moments (see appendix).

## II. Crystal Lattice Distortion

Several studies on o-$R$MnO$_3$ multiferroics have reported a structural distortion within the multiferroic phase [22, 21, 19] (below $T_C$), which is indicative of a lowering of crystal symmetry from Pbnm. One study has suggested that the low temperature symmetry is P2$_1$mn (space group 31), which is polar [23]. The distortion is typically observed using XRD, by studying structural reflections that are forbidden by the high temperature space group. Figure 6 presents the integrated intensity of (033) and (022) as functions of temperature. The (033) reflection is forbidden by Pbnm but allowed in P2$_1$mn, while (022) is allowed in both, making it orders of magnitude more intense, as the structural distortion is small. The magnetically-induced ferroelectric crystal distortion is clearly apparent, as the intensity of (033) increases sharply below $T_C$. A contribution to intensity from the (033) remains above $T_C$, which is the same magnitude as the intensity caused by the magnetically-induced distortion. As the magnetically-induced distortion is known to be very small, we attribute this addition to a very small strain-induced distortion in the film that is independent of temperature. Notably no clear jump is observed below $T_{Ho1} \approx 14.5$ K. Other Pbnm-forbidden reflections were also studied, and none exhibited a signature of $T_{Ho1}$.

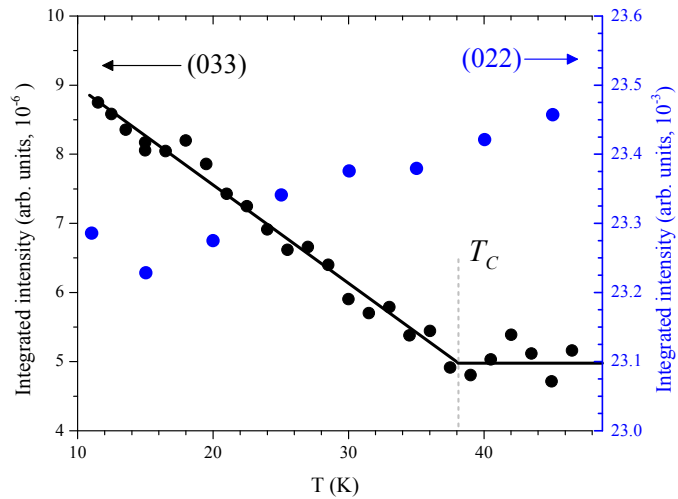

Figure 6 – Temperature dependence of the integrated intensity of structural reflections, taken from a 240nm thick [010]-oriented film. The (033) reflection corresponds to the left axis, the (022) corresponds to the right axis. Integrated intensity values are directly comparable, and all diffracted intensities were corrected for the polarization factor and for variations in the scattering volume due to sample rotation

## Discussion

The most prominent observation in these results is that the Mn AF order is affected by the Ho ordering transition associated with $P$ enhancement ($T_{Ho1}$). A certain level of coupling is to be expected between the Mn and Ho moments. Through dipolar interactions, the Mn order should induce an AF order with the same periodicity onto the paramagnetic Ho moments, as previously shown for $NdNiO_3$ and $o$-$TmMnO_3$ [24, 19]. This is directly evident from the $q$ dependence in Figure 4c. However, unlike these previous reports, we observe clear signatures that Ho $4f$ order affects the Mn $3d$ moments as well. This is primarily noticeable at the $T_{Ho1}$ transition, in Figures 3 and 4. A weak effect is possibly also observed on the Mn intensity around $T_{Ho2} \approx 23$ K.

For clarity, we now limit ourselves to distinguishing between moment components along the $c$ axis ($m_c$) and moments in the $ab$ plane ($m_{ab}$). For the Mn ions, the main magnetic order parameter is described by a $(0\ q\ 1)$ wave vector. This is the order parameter responsible for the lattice distortion, and it is coupled to a second order parameter, described by $(0\ q\ 0)$, which is the order we observe at the Mn $L_3$ edge. The commonly accepted scenario is that the $(0\ q\ 1)$ order is primarily caused by the modulation of the Mn $m_{ab}$, while the $(0\ q\ 0)$ one is caused by the modulation of the Mn $m_c$. The azimuthal dependence of the Mn $(0\ q\ 0)$ reflection is in good agreement with this scenario, and is well-described by the calculation in Figure 5a. This rules out a $bc$ cycloid as the main Mn AF motif (as in the case of $TbMnO_3$), but could agree with an $ab$ cycloid (as in the case of $DyMnO_3$) or a modulated E-type (as suggested in Refs. [21, 19, 25]). The order induced below $T_N$ by the Mn on the paramagnetic Ho moments therefore contains only moment components along the $c$ axis [19]. At the Ho edge, the $(0\ q\ 0)$ and $(0\ q\ 1)$ reflections appear directly below $T_N$, indicating that they are sensitive to the induced order. The $k = 1$-$q$ type reflections do not appear at the Mn $L$ edges, and at the Ho edge they appear, but at temperatures lower than $T_N$. These reflections therefore serve as indicators of additional magnetic transitions that the Ho moments undergo independently, $T_{Ho1}$ and $T_{Ho2}$. They suggest a more complicated structure than that of the Mn, such as the one described in Ref. [6], and are sensitive to this order, not to the induced order from the Mn.

The sensitivity of these reflections to different moment components is described in detail in the appendix: $(0\ k\ 1)$-type reflections are primarily sensitive to $m_{ab}$, while $(0\ k\ 0)$-type are primarily sensitive to $m_c$. The appearance of $(0\ 1$-$q\ 1)$ at $T_{Ho2} = 23$ K therefore suggests an ordering of the Ho $m_{ab}$. Indeed $T_{Ho2}$ bares no clear signature on the $(0\ q\ 0)$ signal at the Ho edge (Fig. 4a), suggesting that the order parameter associated with $T_{Ho2}$ is a weak contribution to this reflection (at $\Psi = 0°$). The appearance of $(0\ 1$-$q\ 0)$ at $T_{Ho1} = 14.5$ K is expected to indicate independent ordering of the Ho $m_c$ components. However, the dominant contribution to the $\Psi$-dependence of $(0\ 1$-$q\ 0)$ in Fig. 5c has a $\pi/2$ phase shift to the c-axis components in Fig. 5a,b. This suggests that despite the preferred sensitivity to $m_c$, the dominant contribution to $(0\ 1$-$q\ 0)$ is an a-axis component. This agrees with the sharp intensity rise observed below $T_{Ho1}$ in the $(0\ q\ 1)$ and $(0\ 1$-$q\ 1)$ reflections (Fig. 4b). An equivalent phase shift was observed in $TbMnO_3$, when measured at the Tb $L$ edges [26, 27]. This strongly suggests that the Ho order below $T_{Ho1}$ is an ac-cycloid-like arrangement, as is the case for ordered Tb moments in $TbMnO_3$ [28, 29]. The evolution of the Ho moments with $T$ is best described by dividing the temperatures below $T_N$ into three ranges, as done in Figure 4. Since the enhancement of **P** occurs only below $T_{Ho1}$ (Figure 1), we associate it with the independent order of the Ho $m_a$. Note that we do not draw a clear conclusion for $m_c$ below $T_{Ho1}$, but suggest that it is ordered.

The mechanism through which the ordered Ho moment relates to the large *P* enhancement is still unclear. The Pbnm-forbidden reflection (033) is sensitive to the magnetically-induced lattice distortion, but showed no change in trend below $T_{Ho1}$. The integrated intensity of such reflections has been previously shown to scale linearly with that of (0 *q* 0) at the Mn edge [21]. This occurs because both are coupled to the main Mn (0 *q* 1) order parameter, as was reported for *o*-LuMnO$_3$ in which *R* ion ordering does not occur. From Figures 4 and 6 it is clear that such a relation with (033) can only hold in the range $T_{Ho2} < T < T_N$. Below this range the Mn signal flattens out with cooling, while the (033) continues to grow linearly. This suggests that the Mn $m_c$ (the (0 *q* 0) order) is not coupled to the lattice distortion occurring due to the influence of the Ho order.

Since the primary Mn order parameter (0 *q* 1) is directly coupled to the Ho $m_c$ moments through dipolar interactions [19], a change in the Ho $m_c$ moment (such as *ac* cycloidal order) could affect or enhance the lattice distortion. Such an enhancement effect of the Ho on the Mn-induced polarization is also supported by results from DyMnO$_3$ and TbMnO$_3$ [30, 31]. In these studies, the *R* ions order, and cause a minor increase in **P**. Furthermore, the *R* ions order with a different *q* than the Mn ions (Tb and Dy order with *q* = 0.415 and 0.5, compared to Mn with 0.27 and 0.36, respectively). This is unlike HoMnO$_3$, in which Ho and Mn keep the same *q* value, indicating a stronger coupling between the two orders. Furthermore, as the Mn order is not a *bc* cycloid (as seen in Fig. 5a, and according to Ref. [19]), one could expect a different response of the Mn-induced polarization, which involves symmetric exchange striction in addition to inverse Dzyaloshinskii-Moriya interaction. However, the absence of any signature of $T_{Ho1}$ in the (033) signal (Figure 6) indicates that the **P** enhancement does not involve this lattice distortion. This could occur if the additional **P** contribution is a purely electronic contribution, which does not involve a further distortion of the lattice. Another consideration is that the (033) is a measure for the strength of the ferroelectric phase, because it appears due to the symmetry lowering caused by the Mn order. However, it does not measure the ferroelectric lattice distortion itself, and may not be coupled to the additional distortion created by the Ho order. This would suggest that the magnetoelectric interaction caused by the Ho moments couples differently to the oxygen cage.

In other *o*-*R*MnO$_3$ materials ordering of *R* ions does affect the lattice. For example, the Tb$^{3+}$ moments in TbMnO$_3$ order at 7 K. This transition, while producing only a weak variation in **P**, has an effect on the phonon spectrum [32]. Shimamoto et al [13] have recently shown that *o*-HoMnO$_3$ films can grow as two layers, strained and relaxed. This provides an alternative explanation for our results. An enhancement of the lattice distortion could have also occurred in our samples of *o*-HoMnO$_3$, if we consider the difference in penetration depth for the two diffraction experiments: the RXD experiment probes magnetic order within the first ~10-20 nm of the film, while the XRD experiment probes the lattice distortion in the entire film. Assuming that the top of the film is relaxed and bulk-like, while the bottom of the film is strained, the $T_{Ho1}$ transition might occur at 14.5 K only in the relaxed part. An enhanced lattice distortion signal from the top few nm would need to be comparable in size to the existing XRD signal from the whole film.

In this case the lattice distortion from $T_{Ho1}$ would most likely not be detected in the XRD experiment in Figure 6, in contrast to the resonantly enhanced magnetic RXD signals.

## Conclusions

Bulk-like samples of thick o-HoMnO$_3$ films have been studied using X-ray diffraction techniques, sensitive either to the Mn AF order, to Ho AF order, or to the ferroelectric lattice distortion. We observe a series of three magnetic transitions. We find that the Mn AF order is strongly affected by an AF ordering transition of Ho moments at 14.5K. The same transition is also associated with a strong enhancement of the ferroelectric polarization, but we find no evidence of an enhancement of the lattice distortion associated with the Mn-induced symmetry-lowering. Based on our results we suggest that the Ho order that forms below 14.5 K is an ac-cycloid, similar to TbMnO$_3$. Ho $m_c$ moments known to couple to the Mn order, but the absence of a Ho-ordering signature on the Mn-induced lattice distortion suggests an alternative route to enhanced polarization which is not through the Mn.

## Acknowledgements

Experiments were performed at the X11MA and X04SA beamlines at the Swiss Light Source, Paul Scherrer Institute (Villigen, Switzerland). We thank Dr. Yi Hu for sample growth at PSI. We thank the X11MA and X04SA beamline staff for experimental support. The financial support of PSI and the Swiss National Science Foundation (SNSF) is gratefully acknowledged, with Y. W. W. through Grants No. 200020-159220 and 200021-137657, M. R. through Sinergia Grant No. CRSII2_147606, and L. R. from the SNSF's National Center of Competence in Research Molecular Ultrafast Science and Technology (NCCR MUST), Grant number 51NF40-183615. We further acknowledge the support of SNSF through project No. 200020-117642 and through the NCCR MaNEP. This work received funding from the DFG within the Emmy Noether program under Grant No. RE 3977/1.

## Appendix

The purpose of this appendix is to gain insight into the sensitivity of the RXD signals to different components of the magnetic moments. To do so, we follow Refs. [33, 34] to employ spherical tensor notation. This is not strictly required, but provides a clearer picture of the observations. In this notation a tensor can be written as $T_Q^K$, in which K is the tensor's rank and Q is its projection. We also limit ourselves to consider dipolar excitations only (E1E1). For a reflection defined by a momentum transfer $\boldsymbol{\tau} = 2\pi(hkl)$, (hkl are miller indices) the resonant structure factor can be written as

$$F(E1, \boldsymbol{\tau}) \propto \sum_{KQ} (-1)^Q \chi_Q^K D_{Qq}^K \Psi_q^K(\boldsymbol{\tau}) \tag{A1}$$

The proportionality term does not concern us in this discussion. Here $X_Q^K$ describes the light, and $\Psi_Q^K$ describes the material ($D_{Qq}^K$ is a rotation of $\Psi_q^K$ to the coordinate system of $\chi_Q^K$). This separation is instrumental, because for understanding the sensitivity of a reflection to specific terms, all that is needed is knowledge of $\Psi_q^K$. This is described as

$$\Psi_Q^K(\boldsymbol{\tau}) = \sum_d \langle T_Q^K \rangle_d e^{i\boldsymbol{\tau}\cdot\boldsymbol{d}} \tag{A2}$$

Here $T_Q^K$ is a tensor of rank K describing a component of the electronic cloud (charge, spin etc…). The brackets denote a time average, and the sum is over all resonant ions in the unit cell. As we concern ourselves only with magnetic dipolar moments, we limit ourselves to a rank of K = 1, and thus projections Q = -1,0,1. To relate cartesian moments to $\langle T_Q^1 \rangle$ objects, one can utilize a simple conversion presented in Ref. [33]: the $c$ axis moment $m_c$ is described by $\langle T_0^1 \rangle$, while the $a$ and $b$ moments are described as $\langle T_{\pm 1}^1 \rangle = (-im_y \mp m_x)/\sqrt{2}$. $\langle T_{\pm 1}^1 \rangle$ are two entities that we will only consider together in this discussion. Alternatively, a more complex object representing a cycloid was used in Ref. [35].

We are now ready to analyze the signals observed. For the Mn ions we use $\boldsymbol{\tau} = 2\pi(0k0)$, since only the (0 $q$ 0) reflection is observed. We plug in the Mn ions' 4b positions into Eq. (A2), and find that $\Psi_{\pm 1}^1 = 0$, while $\Psi_0^1(\boldsymbol{\tau}) = 2\langle T_0^1 \rangle(1 - e^{\pi i k})$. This reaffirms conclusion of previous works, that the (0 $q$ 0) reflection at the Mn edge is sensitive *only* to the $c$ axis moments, i.e. to $\langle T_0^1 \rangle$.

We repeat this for the Ho ions, plugging in their crystallographic 4c positions of type ($x$ $y$ ¼), which are described by two free parameters: $x$ and $y$. For $\boldsymbol{\tau} = 2\pi(0k0)$ we find

$$\Psi_0^1(0k0) = -4i\langle T_0^1 \rangle \cos(2\pi yk) \sin(\tfrac{1}{2}\pi k) \, e^{\tfrac{1}{2}i\pi k}$$
$$\Psi_{\pm 1}^1(0k0) = 2i \sin(2\pi yk) \left( \langle T_{\pm 1}^1 \rangle + \langle T_{\mp 1}^1 \rangle e^{i\pi k} \right) \tag{A3}$$

while for $\boldsymbol{\tau} = 2\pi(0kl)$ we find

$$\Psi_0^1(0kl) = 4\langle T_0^1 \rangle \sin(2\pi yk) \sin(\tfrac{1}{2}\pi k) \, e^{\tfrac{1}{2}i\pi k}$$
$$\Psi_{\pm 1}^1(0kl) = 2\cos(2\pi yk) \left( \langle T_{\pm 1}^1 \rangle + \langle T_{\mp 1}^1 \rangle e^{i\pi k} \right) \tag{A4}$$

For further insight, we inspect the quantity $\zeta = 2\pi yk$. We plug in $k = q \approx 0.40$ and $y \approx 0.084$ [6, 11], and find that $\cos\zeta \approx 1$ while $\sin\zeta \approx 0.2$. This implies that (0 $q$ 0) at the Ho edge is primarily sensitive to $\langle T_0^1 \rangle$ and weakly sensitive to $\langle T_{\pm 1}^1 \rangle$, unlike the Mn resonance, in which sensitivity is *only* to $\langle T_0^1 \rangle$. For (0 $q$ 1) the opposite is true: it is strongly sensitive to $\langle T_{\pm 1}^1 \rangle$ and weakly sensitive to $\langle T_0^1 \rangle$. From these conclusions we conclude that (0 $q$ 1) is primarily sensitive to the Ho $m_{ab}$ and (0 $q$ 0) is primarily sensitive to $m_c$.